# Programmable quantum emitter formation in silicon

K. Jhuria[1, *], V. Ivanov[1], D. Polley[2, 3], W. Liu[1], A. Persaud[1], Y. Zhiyenbayev[2], W. Redjem[2], W. Qarony[2], P. Parajuli[4], Qing Ji[1], A. J. Gonsalves[1], J. Bokor[2], L. Z. Tan[4], B. Kanté[2], and T. Schenkel[1,*]

[1]Accelerator Technology and Applied Physics Division, Lawrence Berkeley National Laboratory, Berkeley, California 94720, USA

[2]Department of Electrical Engineering and Computer Sciences, University of California, Berkeley, California 94720, USA

[3]Department of Physics and Nanotechnology, SRM Institute of Science and Technology, Kattankulathur, Chennai, Tamil Nadu, India

[4]Molecular Foundry, Lawrence Berkeley National Laboratory, Berkeley, CA 94720, USA

(* t_schenkel@lbl.gov, kaushalya@lbl.gov)

## Abstract

Silicon-based quantum emitters are candidates for large-scale qubit integration due to their single-photon emission properties and potential for spin-photon interfaces with long spin coherence times. Here, we demonstrate local writing and erasing of selected light-emitting defects using fs laser pulses in combination with hydrogen-based defect activation and passivation. By selecting forming gas ($N_2/H_2$) during thermal annealing of carbon-implanted silicon, we form $C_i$ centers while passivating the more common G-centers. The $C_i$ center is a telecom S-band emitter with very promising spin properties that consists of a single interstitial carbon atom in the silicon lattice. Density functional theory calculations show that the $C_i$ center brightness is enhanced by several orders of magnitude in the presence of hydrogen. Fs-laser pulses locally affect the passivation or activation of quantum emitters with hydrogen and enable programmable quantum emitter formation in a qubit-by-design paradigm.

## Introduction and Methodology

Silicon (Si)-based quantum emitters are emerging as viable candidates for quantum computing, sensing, networking, and communication owing to bright photon emission in the telecom band, scalability, and ease of integration with both electronics and photonics [1–3]. Prominent emitters have been revisited recently including the W, G, and T centers which involve common elements from standard Si processing in their structure (i.e., hydrogen (H), carbon (C)). Optimization of local (single) center formation is being pursued in process flows that include ion implantation, thermal annealing, and local excitation with focused ion beams and laser pulses [1–8]. Local laser-driven excitation has been optimized for (single) color center formation in high bandgap semiconductors such as diamond, SiC, and boron nitride[9–11], including with in situ feedback for deterministic single center formation. Laser processing of Si has enabled (local) annealing, doping, and defect engineering e. g. for applications in electronic device formation and photovoltaics [4,5,12–14]. Here, we demonstrate programmable defect center formation with local writing and erasing of selected light-emitting defects using fs laser pulses in combination with hydrogen-based defect activation and passivation. We demonstrate this approach with G centers (a pair of two C atoms at substitutional sites paired with the same Si self-interstitial) along with $C_i$ centers (a single C atom at an interstitial site in the silicon lattice). This local processing approach for engineering Si quantum emitters with fs laser pulses paves the way toward large-scale integration of selected quantum emitters and the realization of Si-based quantum networks.

G centers have been realized using standard ion implantation followed by rapid thermal annealing at different temperatures and times in SOI substrates [3,15]. Interestingly, the $C_i$ center, which only involves a (Si-C)$_{Si}$ split-interstitial pair has not received a lot of attention to date [16–19]. We report a simple recipe to form $C_i$ centers using standard ion implantation and rapid thermal annealing. $C_i$ centers form and G-centers are passivated when we replace inert gases (i.e. Ar and $N_2$) with forming gas ($H_2$:10%, $N_2$:90%) (see methods) during the annealing process. Electron paramagnetic resonance (EPR) measurements by Watkins and Brower [18], revealed two optically addressable spin ½ charge states, which qualify the $C_i$ center as a promising spin-photon interface candidate.

## Results

### 1. Formation, writing, and erasing of light emitting centers in SOI with fs laser pulses

An artistic view of writing and erasing of quantum emitters in SOI using fs laser pulses is shown in Fig 1 along with a structural representation of G and $C_i$ centers. Programmable writing and erasing of color centers can be achieved by changing the fs laser fluence. A raster scan presented in the top right corner of Fig 1 shows photoluminescence (PL) emission signals from G and $C_i$ centers after their localized writing and erasing with fs laser pulses. The raster scan was taken with a 1250 nm long pass filter placed before the superconducting nanowire single photon detector (SNSPD) to attenuate the background signal. Fs laser fluence per pulse is increased from the bottom to the top row (with laser fluences at constant levels in each row).

Fig. 2a shows PL spectra at each step in the process flow starting with an as-received SOI after carbon ion implantation and forming gas annealing (pre-processed SOI) resulting in bright $C_i$ center formation. The presence of hydrogen during annealing plays a pivotal role in the formation of $C_i$ centers while passivating the G centers. This is in contrast to many earlier reports, where G-centers dominate emission spectra after carbon ion implantation and thermal annealing in an inert gas ambient [15,19]. Telecom band wavelengths (i.e. 1200-1600 nm) were scanned to detect the resulting color centers, the selected range is shown for better visualization. Generally, it has been observed that the linewidth of quantum emitters broadens when transitioning from bulk Si to SOI due to the $Si/SiO_2$ interface-induced stress and strain [20]. Interestingly, the $C_i$ centers formed with the above-mentioned recipe provide an extremely narrow linewidth of ~0.03 nm (4.2GHz), a measurement value that is limited by the spectrometer resolution (see supplementary section 3 for a high-resolution spectrum). This can be attributed to the diffusion of H into Si during the annealing process and the formation of H clusters and trapping of impurities to compensate for strain at the $Si/SiO_2$ interfaces [21]. A PL saturation curve was also recorded for $C_i$ centers along with the temperature response to check its stability and robustness (see supplementary section 2).

The pre-processed SOI was then subjected to fs laser pulses of varying fluences in the range ~16-48 $mJ/cm^2$. This fluence range is ~4 times lower than the damage threshold of Si for fs-laser pulses [22,23]. Hence the silicon lattice is not damaged, but fs-pulses in this fluence range can act on the H-bonds to atoms and defect structures in the silicon matrix. Hence, fs-laser pulses enable us to write and erase quantum emitters locally. The last three curves in Figure 2 show the PL spectra obtained after fs-laser irradiation at different laser fluences demonstrating first the writing of G centers along with modifying the density of pre-existing $C_i$ centers followed by partial and complete passivation of G and $C_i$ centers, respectively, at higher laser fluences. The final curve presents the reoccurrence of G and $C_i$ centers after irradiation with a higher laser fluence pulse of ~44.5 mJ/cm$^2$. The PL emission peak intensity corresponding to G and $C_i$ centers is shown in Fig. 2b before and after fs irradiation indicating the writing and erasing of these emitters. Spectra corresponding to the intermediate laser fluences can be found in the supplementary information section 4. While the in-depth understanding of the mechanism governing the writing and erasing of $C_i$ centers with single fs laser irradiation requires further exploration, it can be essentially understood as multi-photon absorption leading to energy transfer to the electronic system, which can be further transferred to the silicon lattice, that in turn couples to the hydrogen atoms bonded to carbon and silicon atoms in defect complexes. Fs-laser fluences well below the silicon damage threshold enable defect center reconstruction into optically dark states and the removal of hydrogen from $C_i$ centers. Increasing the laser fluence to ~40 mJ/cm2 can aid the redistribution of H and leads to the re-activation of $C_i$ centers [24]. Writing of G centers can be understood in a similar way where the breaking of an H-bond in the vicinity of an optically inactive G center (A-configuration) may lead to the formation of an optically active G-center in the B configuration [25]. Time-resolved photoluminescence (TR-PL) measurements were also performed before and after the fs irradiation to extract the non-radiative lifetimes for both G and $C_i$ centers. The lifetime of $C_i$ centers was found to be ~3 ns before the fs laser irradiation and changed between 3 ns and 8 ns for different laser fluences. A similar lifetime range was also found for the G centers after fs irradiation. It can be noted that a slight change in the fs laser fluence can change the lifetime of the quantum emitters and hence provides fine control over the quality of the emitters.

## 2. First-principles calculations of $C_i$ centers

We performed first-principles calculations of $C_i$ centers, using a computational workflow described in earlier publications [26], with atomic defect structure corresponding to a split (Si-C)$_{Si}$ pair sharing a substitutional site (Fig. 3a) [16] (see methods). The energy level diagram for the neutral charge state of the C$_i$ center is shown in Fig. 3e, while the diagrams for other charge states and a plot of the chemical potential-dependent formation energies can be found in Supplementary Information section 5.1. We find that the 0, -1, -2, and -3 charge states of the C$_i$ center are stable within the Si gap, with the -1 charge state being stable in intrinsic Si. This is in contrast to prior work [16], which found the +2, +1, and 0 charge states to be stable, albeit with a different correction scheme and a smaller supercell size, which leads to much larger errors due to periodic image charges. As the -3 charge state is only stable in heavily n-doped Si, and the -2 charge state has all defect levels occupied, we focus on the -1 and 0 charge states, as well as the +1 charge state due to it being found stable in prior work.

The computed zero-phonon lines and transition dipole moments of the -1, 0, +1 states are shown in Table 1. As the number of electrons in the center changes, the energies of newly occupied defect levels shift, but the localized defect wavefunctions remain relatively unchanged, localizing to the carbon atom for the lower defect levels, and to the Si for the upper defect levels. The optical properties of these defects can be considered in several ways. First, the Kohn-Sham energy differences (ΔKS) are 968 meV, 817 meV, and 812 meV for the -1, 0 and +1 charge states respectively, and are known to be a decent estimate for the zero-phonon line (ZPL) of defects, deviated by ~100-150 meV [26, 27]. The computed ZPLs for the -1, 0, and +1 charge states using the constrained occupation method are 571 meV, 569 meV, 568 meV respectively. Due to the finite size effects of the supercell and a limited number of k-points, these values can also deviate from the true ZPL by several tenths of an eV [25,26]. Given these considerations, the different charge states of this split (Si-C)$_{Si}$ pair structure may explain the multiple experimentally observed ZPL peaks. However, it should be noted that the described transitions have vanishingly small transition dipole moments, suggesting that this structure might be optically dark.

The small transition dipole moment of the C$_i$ center is the result of the two mirror plane symmetries of the defect and the localization of defect states to different atoms in the defect. This is analogous to the G center, where the A configuration with an interstitial carbon atom is relatively optically dark [25]. By distorting the C$_i$ center so that the carbon occupies the substitutional position, which we term the "B" configuration in analogy to the optically bright G center B, the brightness of the C$_i$ center is enhanced considerably (Table 1), though this structure is 0.66 eV higher in energy. If the mirror symmetries are broken by displacing atoms of the $C_i$ center out of the plane, they are restored during the force relaxation step, suggesting this is not a stable configuration.

Exposure to forming gas during thermal annealing introduced hydrogen into the silicon lattice, leading to the possibility of generating H-related centers. Inspired by the Si T center, where two carbon atoms occupying a substitutional site are bound to an H atom[2], we consider modified $C_i$ center defects, where a hydrogen atom is either bound to the carbon atom in a planar configuration ($C_i$+H Type 1), bound to the carbon with an additional out of plane distortion ($C_i$+H Type 2, Fig. 3c), or bound to the Si atom ($C_i$+H Type 3, Fig. 3d). These modified defects have slightly shifted ZPLs, and are all significantly optically brighter than the bare $C_i$ center (Table 1). They further have a spin-1/2 degree of freedom in the neutral charge state, like the T center.

Aside from the known ZPL peak at 1448 nm (856 meV), a number of other peaks at 1415.4 nm, 1441.7 nm, 1444.3 nm, 1450.8 nm, and 1453.6 nm can be seen in the experimental PL spectra, which correspond to respective shifts of +20 meV, +4 meV, +2 meV, -2 meV, and -3 meV in energy (see supplementary section 3). Our first-principles calculations show that there are shifts in the ZPL of the same magnitude arising from charge states, structural changes, and the presence of H (last column, Table 1). Considering that the deviation of the computed 569 meV ZPL from the observed 856 meV ZPL for the lone $C_i$ center due to systematic errors of the finite unit cell size and constrained occupation method is not strongly structure dependent, we infer that variations in the charge and structure of the $C_i$ center in the presence of hydrogen can explain the variety of observed bright ZPL peaks.

3. **All-Optical writing and erasing of W and G centers in SOI**

While the G and $C_i$ centers were programmed via fs laser pulse irradiation in pre-processed SOI, direct writing and erasing of G centers were also established in as-received SOI without any external ion implantation and thermal annealing. We repeated a similar single fs laser pulse irradiation on the as-received SOI and checked the emission spectra in the telecom band range. Precise control over the G centers was observed in the lower laser fluence regime followed by W centers writing at the near damage threshold fluence range.

Fig. 4a shows the PL spectrum corresponding to G centers ZPL and phonon sidebands after a single fs pulse irradiation at a laser fluence of 28.35 mJ/cm$^2$. The observed trends in the PL peak intensity around this fluence are shown in Fig. 4b. It is clear from the trend that the density of G centers can be precisely controlled by varying the laser fluence.

A recent study using FIB to form G centers demonstrated the necessity of C pre-implantation, without which subsequent processing with a Si-focused ion beam would lead to the formation of only W centers. Our current work with fs laser pulses provides evidence for direct writing and precise control over the density of G center formation on the as-received SOI sample [4]. Additionally, no W centers were observed while working with fs laser pulses in the lower fluence range, further suggesting that this approach is superior for maintaining the integrity of the Si lattice. Even though this article focuses on ensembles of G and $C_i$ centers, control at the single center level can also be achieved with in-situ feedback, as has been demonstrated already for the formation of e. g. nitrogen-vacancy (NV) centers in diamond [28]. Fig. 4c shows the PL spectra after the as-received SOI was irradiated with near Si damage threshold fs laser pulse fluence of ~300 mJ/cm$^2$. W centers were formed near these high fluences consistent with their structure consisting of three silicon interstitials that are formed by damaging the silicon lattice. We performed TR-PL measurements to extract the non-radiative lifetime and to gain insights into the defect dynamics involved in emitter formation and the resulting quality of G centers. Fig. 4d shows the non-radiative lifetime and linewidth of G centers after fs irradiation at different laser fluences. An inverse trend for lifetime and linewidth was observed as a function of fs laser fluence.

## Discussion

**Formation mechanism of G and $C_i$ centers**

The G center in Si is among the most common defect centers and can be formed readily by C ion implantation followed by rapid thermal annealing under inert gas ambiance, as well as by proton irradiation, and laser-ion doping [29,30]. The formation of the G center follows a widely accepted two-step process. Firstly, radiation damage leads to the creation of Si self-interstitial and an accompanying $C_i$ due to ion implantation. The $C_i$, known for its high mobility, can freely migrate within the lattice until it becomes trapped by a substitutional carbon atom, $C_s$. This trapping event results in the formation of the optically inactive A configuration of the G center. However, this configuration can overcome a small potential barrier (~0.14 eV) and undergo a structural transformation to become the optically active B configuration. The B configuration comprises two C atoms located at substitutional sites, both bonded to the same Si self-interstitial[25].

**Passivation of G center with forming gas annealing**

Hydrogen is widely used for defect passivation in semiconductor processing, e. g. to increase minority carrier lifetimes and to reduce interface charge densities[31]. Hydrogen has been previously used to passivate many other defect centers in Si [32], and this is further demonstrated here in the effect of H on the G center. In the optically active B form of the G center, the unpaired electrons are localized to the self-interstitial Si atom and can trap H atoms. We consider two configurations for the H, in-plane, and out-of-plane (See Supplemental Information section 5.2), computing the dipole moments for transition associated with the bright G center luminescence, from the valence band maximum to the midgap localized defect level [25]. The squared transition diploe moments (TDM) for these transitions are 0.067 Debye$^2$ and 0.0845 Debye$^2$, significantly less than the ~5 Debye$^2$ of the native G center, which can explain why G centers are not observed in the PL data from SOI samples that had been preprocessed with thermal annealing in a forming gas ambient shown above (1$^{st}$ curve from Fig. 2a).  Complementary to the G center, the TDMs of the Ci center with additional hydrogen increase by several orders of magnitude (Table 1).

**Writing and erasing mechanism of quantum emitters in Si under fs irradiation**

While the writing and erasing of G and C$_i$ centers with fs laser in our pre-processed SOI samples could be understood as the fs-laser pulses acting on H-bonds, more complicated physical mechanisms are also possible, where the Si-Si and Si-C pair get disassociated due to energy deposition and electronic excitation effects with direct fs laser irradiation, leading to a rearrangement of impurities and dopant atoms in the silicon lattice.  Fs-laser pulses of varying intensity can steer these re-configuration processes and hence enable the writing and erasing of selected quantum emitters with properties optimized for selected applications, from quantum communications to quantum sensing in a qubit by design paradigm.  As a topic for future research, an in-depth understanding of these mechanisms will require simulations of non-adiabatic dynamics using methods such as time-dependent density functional theory, which has shown promise in the study of defect formation in the excited state [33–36].

**Conclusions**

In this article, we address challenges in the development of single photon emitters and spin-photon interfaces, i.e. programmable local formation of qubit candidates with tailored properties. We demonstrate the writing, erasing, and rewriting of quantum emitters in Si using fs-laser pulses. While the approach can be generalized to other emitters in Si, we have demonstrated it on the most common and widely studied light emitting defects in silicon, the W, G centers, together with the re-discovered $C_i$ center in SOI wafers. The optical properties of the $C_i$ center qualify it as a highly promising candidate for applications as a single photon source and potential spin-photon interface due to its relatively simple structure, bright emission in the telecom S-band, relatively narrow spectral linewidth, robustness, and spin degree of freedom in its ground state. Density function theory calculations highlight the role of hydrogen in boosting the brightness of $C_i$ centers with hydrogen. The quantum emitters formed by fs-laser irradiation are ensembles at this stage, but single centers can be selected in samples with low center densities and single centers can be formed through adaptation of in-situ feedback techniques by combining the fs laser with a PL characterization setup. This new approach for deterministic programming of desired quantum emitters paves the way towards scalable quantum networks [and engineering of qubits by design.

**Methods**

**Ion implantation and Rapid thermal annealing**

Samples cut from commercial SOI wafers (220 nm thick device layer, 10-20 Ohm cm, p-type, 2 μm $SiO_2$ box) were used for this study. For the pre-processed SOI, the as-received SOI was first implanted with 38 keV carbon ions ($^{13}C$) targeting a mean implantation depth of ~115 nm (i.e. center of the 220 nm Si device layer). $^{13}C$ implanted SOI was treated by rapid thermal annealing in forming gas (10% $H_2$, 90%$N_2$) at 800 °C for 120 seconds. RTA in forming gas leads to the repair of the implant damage, passivation of G-centers, and the formation of bright $C_i$ centers. The $^{13}C$ isotope of carbon was chosen to enable differentiation of the ion-implanted carbon from any $^{12}C$ present that was present in the starting material (e. g. using secondary ion mass spectrometry, SIMS). $^{13}$C also enables the exploration of quantum memory in the nuclear spin state [37]. (see supplementary section 1 for further details on ion implantation and rapid thermal annealing along with depth-resolved SIMS profile of common elements in SOI before and after ion implantation followed by forming gas thermal annealing).

**Fs laser irradiation**

An amplified Ti:sapphire laser, working at a repetition rate of 250 kHz, with a wavelength centered around 800 nm and a pulse length of 90 fs (full width at half maxima, FWHM) was used for the experiment in single-shot mode. The pulse duration was measured via auto-correlation with an APE Pulse link 150. We used a laser spot size of 20×20 µm$^2$ for local defect center processing. The spot size was measured directly with a beam profilometer at sample position [38]. Samples were irradiated using single fs laser pulses with varied energy per pulse ranging from ~65-195 nJ (corresponding to a laser fluence range of ~16-48 mJ/cm$^2$) to form and passivate G and the $C_i$ quantum emitters. For the formation of W centers, we increased the laser fluence up to ~300 mJ/cm$^2$.

**Characterization**

Photoluminescence (PL) and time-resolved photoluminescence (TR-PL) measurements were performed at 6 K using a scanning confocal microscope for near-infrared spectroscopy. A 532 nm continuous wave and pulsed laser focused onto the sample via a high numerical-aperture microscope objective (NA= 0.85) was used for the optical excitation. The same objective was directed to a spectrometer coupled to an InGaAs camera (900-1620 nm at -80 C). The telecom band wavelengths ranging from 1200-1600 nm were scanned with a grating of 150 g/mm in front of the camera. The absorption depth of 1 µm in Si with a 532 nm laser sets the probing depth for the quantum emitters. Detailed information on the optical setup used for characterization can be found in Y. Zhiyenbayev *et. al.*, 2023 [15].

**First-principles DFT calculations**

In our calculations, a given defect was embedded into a 3 x 3 x 3 supercell of Si, and VASP [39,40] was used to completely relax the atomic positions and obtain the electronic structure. For the electronic structure, the HSE06 [41] functional was used, with the following parameters: a 450 eV energy cutoff, convergence tolerance of 10$^{-10}$ eV, and a force tolerance of 0.001 eV/Å. Excited state energies were computed using the constrained occupation method [42,25], with the same parameters, while post-processing and real space wavefunctions were extracted using VASPKIT [43]. Formation energies were extracted from calculations using PBE functionals[44], and corrected for finite size effects using the Spinney package [45, 46]. All calculations were performed at the Γ-point.

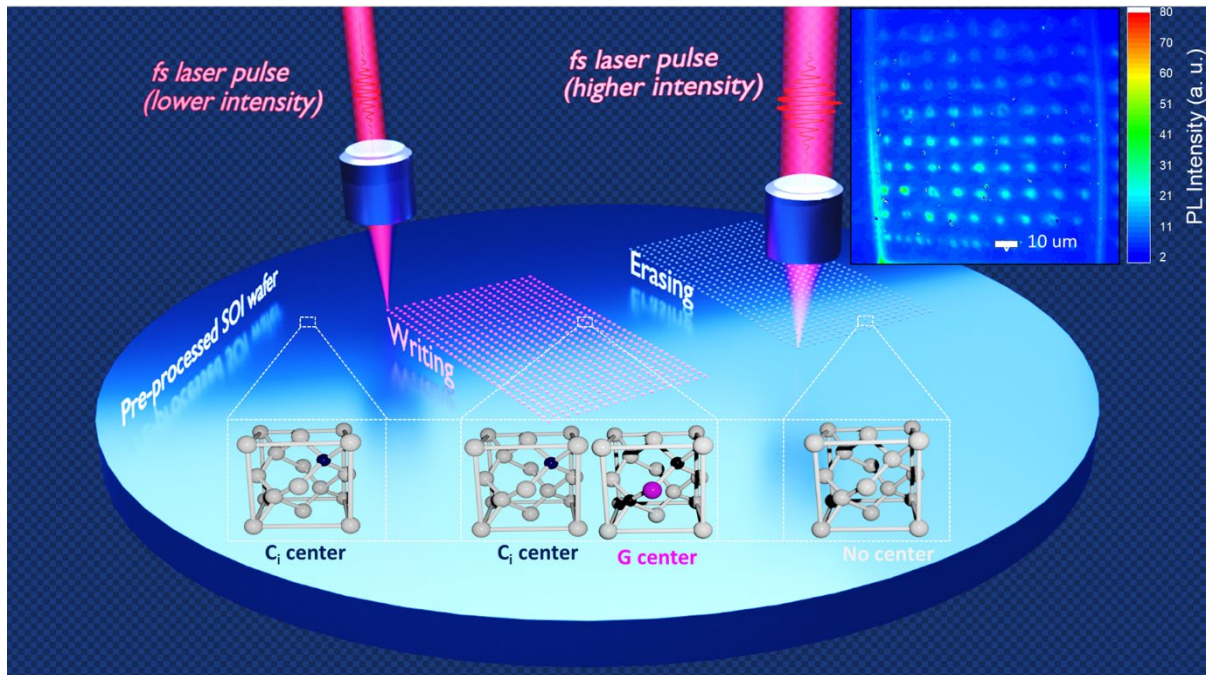

**Fig 1. Programmable quantum emitter formation with fs laser pulses in silicon-on-insulator (SOI).** Artistic representation of the fs laser irradiation approach to write and erase G, and $C_i$ centers in SOI. The G center is a pair of two carbon atoms at substitutional sites (black sphere) combined with the same Si self-interstitial (pink sphere), whereas the $C_i$ consists of a pair of an interstitial carbon (green sphere) and a substitutional Si atom (gray sphere) in the Si lattice (left inset). A single fs laser pulse (90 fs), with wavelength centered at 800nm, was used for irradiation at varied fluences in order to form and erase quantum emitters. Three different irradiation areas are shown to represent the workflow starting with pre-processed SOI wafer with $C_i$ centers formed after ion implantation and rapid thermal annealing under forming ambiance (area on the right). The second area (in the middle) represents the writing of G centers along with modified $C_i$ centers on the pre-processed SOI sample via single fs laser pulse irradiation at relatively low fluences (<30 mJ/cm$^2$). The third area (on the left) shows the erasing of quantum emitters after irradiation with a higher fluence fs laser pulse. A photoluminescence raster scan presenting fs irradiation spots on the pre-processed SOI sample is shown in the top-right corner. Laser Fluence per pulse is increased from bottom to top row (and the laser fluence was held constant in each row).

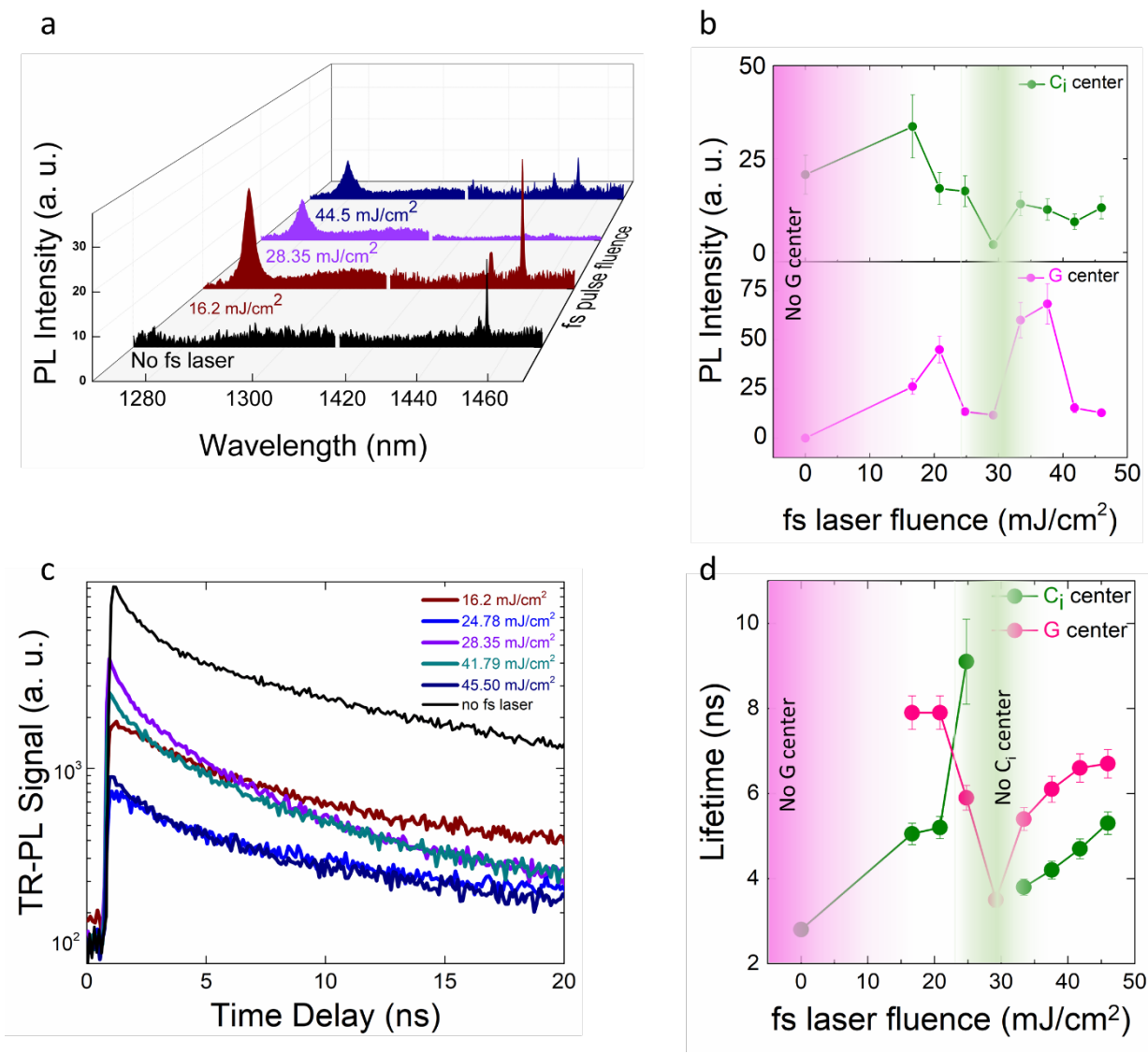

**Fig 2. Writing and erasing of G and $C_i$ center with fs laser pulses below the damage threshold of Si. a.** Process flow represented by measuring the PL spectra starting from the pre-processed SOI sample (emission ~1453 nm, corresponding to the $C_i$ center after carbon ion implantation (7e13 C/cm$^2$ fluence) followed by rapid thermal annealing at 800°C for 120 seconds under forming gas ambiance). The last three curves show the PL spectra obtained after fs irradiation at different laser fluences in order to first write G centers along with modifying the density of pre-existing $C_i$ centers (~16 mJ/cm$^2$), followed by partial and complete passivation of G and $C_i$ centers respectively at higher fluences (~30 mJ/cm$^2$). The final curve presents the reoccurrence of G and $C_i$ centers after irradiation with an even higher fluence pulse (i.e., 44.5 mJ/cm$^2$). The damage threshold in our experiments was >100 mJ/cm$^2$. **b.** PL emission peak intensity of G and $C_i$ centers after irradiation with fs pulses of varying energies. **c.** TR-PL signal from the $C_i$ centers before and after the fs laser irradiation to extract non-radiative lifetimes. d. Lifetimes as a function of fs laser fluence for both G and $C_i$ centers extracted by fitting the TR-PL signal with a first-order decay function.

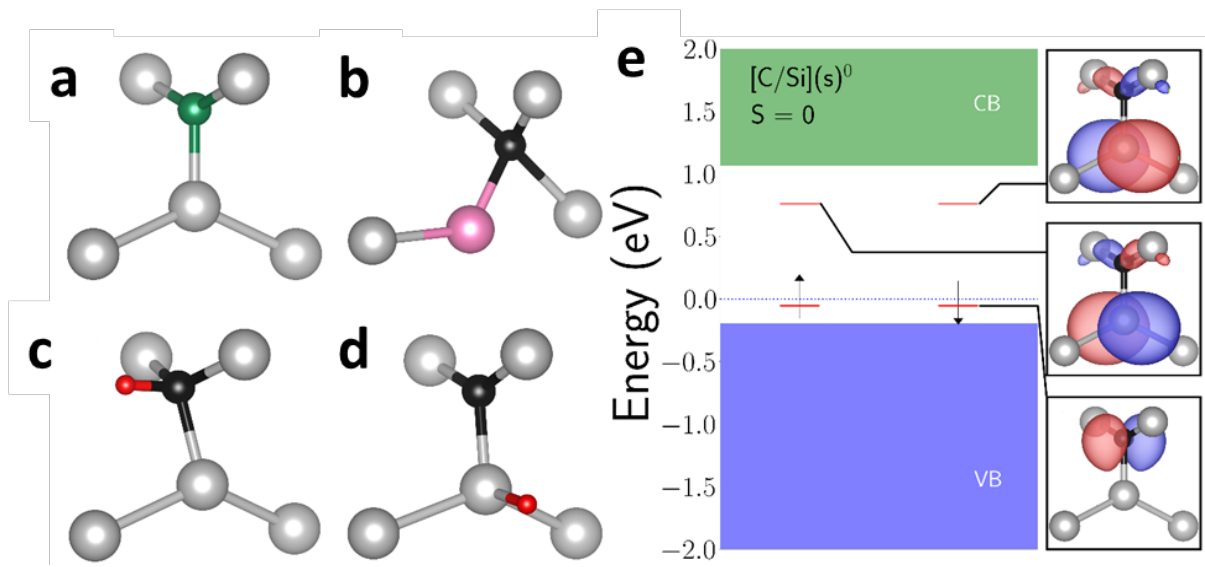

**Fig. 3 Defect levels, structures, and modifications of the C$_i$ center.** Structure of **a.** the C$_i$ center, **b.** the displaced "B" configuration of the C$_i$ center, and modified versions of the C$_i$ center with H bonded to either **c.** the carbon, or **d.** the Si atom. Atoms are colored as follows – Si (grey), carbon (black, green), H (red), Si self-interstitial (pink). **e.** Energy level diagrams for the neutral charge state of the C$_i$ center, with conduction band (green), valence band (blue), localized defect levels (red), and electron occupation (black arrows) indicated. Panels on the right show the real space wavefunctions corresponding to each localized defect level.

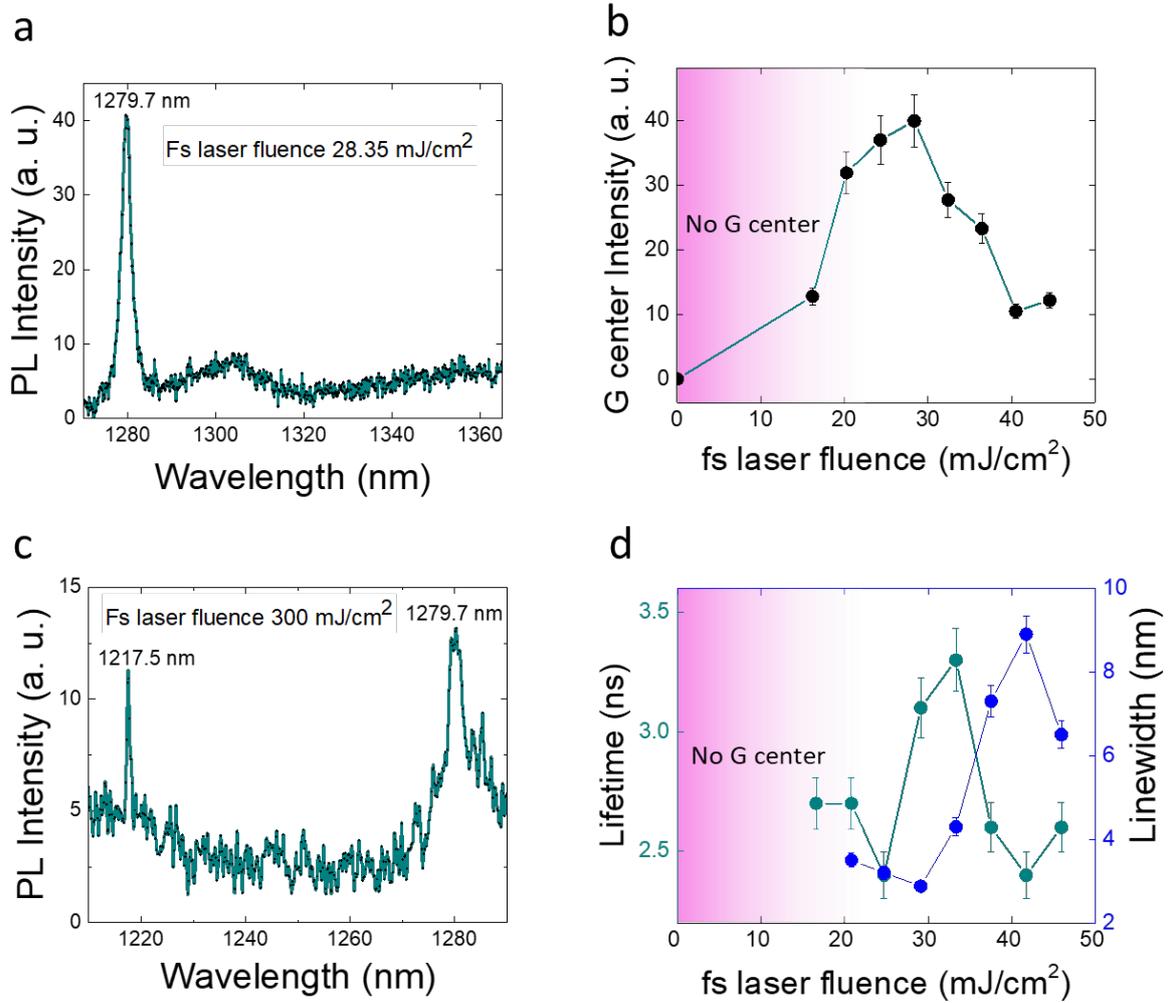

**Fig. 4 All-optical writing and erasing of G centers with direct fs laser pulse irradiation on as-received SOI. a.** PL response from G centers showing the ZPL and the corresponding phonon side band formed in as-received SOI substrate after direct irradiation with a single fs laser pulse of 28.35 mJ/cm$^2$ fluence. **b.** The peak amplitude of the PL spectra from G centers as a function of fs laser fluence per pulse represents the writing and partial erasing of these emitters with fine control over the density. **c.** PL spectra after fs irradiation with near damage threshold fluence of ~300 mJ/cm$^2$ leading to the formation of radiation damage related W centers along with G centers. **d.** The extracted lifetime of G centers after fitting the TR-PL signal with a single decay function along with their linewidths after fs irradiation for a series of laser fluences.

| Defect | Zero-Phonon Line (ZPL) (meV) | Spin Channel | Squared Transition Dipole Moment (TDM) (Debye$^2$) | Relative Shift (meV) |
|---|---|---|---|---|
| $C_i$ center (-1) | 571 | Down | 0.169 x 10$^{-5}$ | +2 |
| $C_i$ center (0) | 569 | N/A | 0.337 x 10$^{-5}$ | |
| $C_i$ center (+1) | 568 | Down | 0.482 x 10$^{-8}$ | -1 |
| $C_i$ center "B" (0) | 655 | N/A | 0.118 | +86 |
| $C_i$+H Type 1 (0) | 103$^2$ | Down | 0.29 x 10$^{-2}$ | +463 |
| | 528 | Up | 0.483 | -41 |
| $C_i$+H Type 1 (+1) | 611 | N/A | 0.632 | +42 |
| $C_i$+H Type 2 (0) | 847 | Down | 0.28 x 10$^{-2}$ | +278 |
| | 559 | Up | 0.721 | -10 |
| $C_i$+H Type 2 (+1) | 581 | N/A | 0.633 | +12 |
| $C_i$+H Type 3 (0) | 1174 | Down | 2.52 | +578 |
| | 592 | Up | 0.902 x 10$^{-4}$ | +23 |
| $C_i$+H Type 3 (+1) | 591 | N/A | 2.20 | +22 |

**Table. 1 Energies and dipole moments of $C_i$ center charge states and modifications.** Computed ZPL and squared transition dipole moment (TDM) for different charge states and structural modifications of the $C_i$ center in Si. For defects with a spin degree of freedom, the spin channel for the transition is indicated. The final column gives the ZPL deviation for each defect relative to the neutral $C_i$ center.

## Authors contribution

T.S. and K.J. conceived and designed the experiments. K.J. and D.P. performed the single shot fs irradiation experiments with guidance from T.S. and J.B. K.J. performed the PL experiments with help from Y.Z. and W.R., with guidance from T.S. and B.K. K.J., V.I., and T.S. analyzed the results with inputs from all authors. V.I. and P.P. performed the first principles-based calculations and summarized the theoretical results in the manuscript with guidance from L.Z.T. K.J. and T.S. wrote the manuscript with inputs from all authors.

## Acknowledgments

This work was supported by the Office of Science, Office of Fusion Energy Sciences, of the U.S. Department of Energy, under Contract No. DE-AC02-05CH11231. L. Z. T and V. I. were supported by the Molecular Foundry, a DOE Office of Science User Facility supported by the Office of Science of the U.S. Department of Energy under Contract No. DE-AC02-05CH11231. This research used resources of the National Energy Research Scientific Computing Center, a DOE Office of Science User Facility supported by the Office of Science of the U.S. Department of Energy under Contract No. DE-AC02-05CH11231.

# References

1. Baron, Y. *et al.* Detection of Single W-Centers in Silicon. *ACS Photonics* **9**, 2337–2345 (2022).
2. Higginbottom, D. B. *et al.* Optical observation of single spins in silicon. *Nature* **607**, 266–270 (2022).
3. Redjem, W. *et al.* Single artificial atoms in silicon emitting at telecom wavelengths. *Nature Electronics 2020 3:12* **3**, 738–743 (2020).
4. Quard, H. *et al.* Femtosecond laser induced creation of G and W-centers in silicon-on-insulator substrates. (2023). *arXiv: 2304.03551*
5. Andrini, G. *et al. Efficient activation of telecom emitters in silicon upon ns pulsed laser annealing.* arXiv:2304.10132.
6. Hollenbach, M. *et al.* Wafer-scale nanofabrication of telecom single-photon emitters in silicon. *Nat Commun* **13**, (2022).
7. Skolnick, M. S., Cullis, A. G. & Webber, H. C. Defect-Induced Photoluminescence from Pulsed Laser Annealed Si. *MRS Online Proceedings Library* **1**, (1980).
8. Komza, L. *et al.* Indistinguishable photons from an artificial atom in silicon photonics. (2022). *arXiv: 2211.09305*
9. Gao, X. *et al.* Femtosecond Laser Writing of Spin Defects in Hexagonal Boron Nitride. *ACS Photonics* **8**, 994–1000 (2021).
10. Ishmael, S. N. *et al.* Laser writing of individual nitrogen-vacancy defects in diamond with near-unity yield. *Optica, Vol. 6, Issue 5, pp. 662-667* **6**, 662–667 (2019).
11. Chen, Y. C. *et al.* Laser Writing of Scalable Single Color Centers in Silicon Carbide. *Nano Lett* **19**, 2377–2383 (2019).
12. Fortunato, G., Mariucci, L., Pecora, A., Privitera, V. & Simon, F. Historical evolution of pulsed laser annealing for semiconductor processing. *Laser Annealing Processes in Semiconductor Technology: Theory, Modeling and Applications in Nanoelectronics* 1–48 (2021) doi:10.1016/B978-0-12-820255-5.00006-4.
13. Sun, Z. & Gupta, M. C. Laser induced defects in silicon solar cells and laser annealing. *Conference Record of the IEEE Photovoltaic Specialists Conference* **2016-November**, 713–716 (2016).
14. Poate, J. M. & Brown, W. L. Laser annealing of silicon. *Phys Today* **35**, 24–30 (1982).
15. Zhiyenbayev, Y. *et al.* Scalable manufacturing of quantum light emitters in silicon under rapid thermal annealing. *Opt Express* **31**, 8352 (2023).
16. Wang, H., Chroneos, A., Londos, C. A., Sgourou, E. N. & Schwingenschlögl, U. Carbon related defects in irradiated silicon revisited. *Sci Rep* **4**, (2014).
17. Capaz, R. B. *et al. Identification of the migration path of interstitial carbon in silicon. PHYSICAL REVIEW B* vol. 50 (1994).
18. Watkins, G. D. & Brower, K. L. EPR observation of the isolated interstitial carbon atom in silicon. *Phys Rev Lett* **36**, 1329–1332 (1976).
19. Durand, A. *et al.* Broad diversity of near-infrared single-photon emitters in silicon. (2020) doi:10.1103/PhysRevLett.126.083602.
20. Camassel, J., Falkovsky, L. A. & Planes, N. Strain effect in silicon-on-insulator materials: Investigation with optical phonons. *Phys Rev B* **63**, 035309 (2000).
21. Sopori, B., Zhang, Y. & Ravindra, N. M. Silicon Device Processing in H-Ambients: H-Diffusion Mechanisms and Influence on Electronic Properties. *J Electron Mater* **30**, (2001).
22. Sladek, J. & Mirza, I. M. Laser induced damage threshold of silicon with native and artificial SiO2 layer. *MM Science Journal* **2019**, 3579–3584 (2019).

23. Bonse, J., Baudach, S., Krüger, J., Kautek, W. & Lenzner, M. Femtosecond laser ablation of silicon-modification thresholds and morphology. *Appl Phys A Mater Sci Process* **74**, 19–25 (2002).
24. Leary, P. & Jones, R. *Interaction of hydrogen with substitutional and interstitial carbon defects in silicon*. *Phys Rev B* **7**, 57 (1998).
25. Ivanov, V. *et al.* Effect of localization on photoluminescence and zero-field splitting of silicon color centers. *Phys Rev B* **106**, 134107 (2022).
26. Ivanov, V. *et al.* Database of semiconductor point-defect properties for applications in quantum technologies. (2023). *arXiv: 2303.16283*
27. Xiong, Y. *et al.* High-throughput identification of spin-photon interfaces in silicon. (2023). *arXiv: 2303.01594*
28. Chen, Y.-C. *et al.* Laser writing of individual nitrogen-vacancy defects in diamond with near-unity yield. *Optica* **6**, 662 (2018).
29. Liu, W. *et al.* Quantum emitter formation dynamics and probing of radiation induced atomic disorder in silicon. (2023), *Phys. Rev. Applied* (in press), *arXiv: 2302.05814*
30. Redjem, W., Amsellem, A.J., Allen, F.I. *et al.* Defect engineering of silicon with ion pulses from laser acceleration. *Commun Mater* **4**, 22 (2023).
31. Pearton, S. J., Corbett, J. W. & Shi, T. S. *Hydrogen in Crystalline Semiconductors*. *Appl. Phys. A* vol. 43 (1987).
32. Sveinbjörnsson, E. Ö. & Engström, O. Hydrogen passivation of gold centers in silicon. *Materials Science and Engineering B* **36**, 192–195 (1996).
33. Apostolova, T., Kurylo, V. & Gnilitskyi, I. Ultrafast Laser Processing of Diamond Materials: A Review. *Front Phys* **9**, 306 (2021).
34. Shimotsuma, Y. *et al.* Formation of NV centers in diamond by a femtosecond laser single pulse. *Opt Express* **31**, 1594 (2023).
35. Liu, Y. Y. *et al.* Electronically induced defect creation at semiconductor/oxide interface revealed by time-dependent density functional theory. *Phys Rev B* **104**, 115310 (2021).
36. Lee, C. W., Stewart, J. A., Dingreville, R., Foiles, S. M. & Schleife, A. Multiscale simulations of electron and ion dynamics in self-irradiated silicon. *Phys Rev B* **102**, (2020).
37. Balasubramanian, G. *et al.* Ultralong spin coherence time in isotopically engineered diamond. *Nat Mater* **8**, 383–387 (2009).
38. Magnes, J. *et al.* arXiv:physics/0605102v1 [physics.optics] *Quantitative and Qualitative Study of Gaussian Beam Visualization Techniques*. (2006).
39. Phys. Rev. B 47, 558(R) (1993) - Ab initio molecular dynamics for liquid metals. https://journals.aps.org/prb/abstract/10.1103/PhysRevB.47.558.
40. Kresse, G. & Joubert, D. From ultrasoft pseudopotentials to the projector augmented-wave method. *Phys Rev B* **59**, 1758 (1999).
41. Krukau, A. V., Vydrov, O. A., Izmaylov, A. F. & Scuseria, G. E. Influence of the exchange screening parameter on the performance of screened hybrid functionals. *Journal of Chemical Physics* **125**, (2006).
42. Gali, A., Janzén, E., Deák, P., Kresse, G. & Kaxiras, E. Theory of Spin-Conserving Excitation of the N-V-Center in Diamond. *Phys Rev Lett* **103**, (2009).
43. Wang, V., Xu, N., Liu, J. C., Tang, G. & Geng, W. T. VASPKIT: A user-friendly interface facilitating high-throughput computing and analysis using VASP code. *Comput Phys Commun* **267**, 108033 (2021).
44. Perdew, J. P., Burke, K. & Ernzerhof, M. Generalized gradient approximation made simple. *Phys Rev Lett* **77**, 3865–3868 (1996).

45. Arrigoni, M. & Madsen, G. K. H. Spinney: Post-processing of first-principles calculations of point defects in semiconductors with Python. *Comput Phys Commun* **264**, 107946 (2021).
46. Kumagai, Y. & Oba, F. Electrostatics-based finite-size corrections for first-principles point defect calculations. *Phys Rev B Condens Matter Mater Phys* **89**, (2014).
47. Redjem, W. *et al.* All-silicon quantum light source by embedding an atomic emissive center in a nanophotonic cavity. *Nature Communications 2023 14:1* **14**, 1–7 (2023).

Supplementary information for

# Programmable quantum emitter formation in silicon

K. Jhuria[1, *], V. Ivanov[1], D. Polley[2,3], W. Liu[1], A. Persaud[1], Y. Zhiyenbayev[2], W. Redjem[2], W. Qarony[2], P. Parajuli[4], Qing Ji[1], A. J. Gonsalves[1], J. Bokor[2], L. Z. Tan[4], B. Kanté[2], and T. Schenkel[1]

[1]Accelerator Technology and Applied Physics Division, Lawrence Berkeley National Laboratory, Berkeley, California 94720, USA

[2]Department of Electrical Engineering and Computer Sciences, University of California, Berkeley, California 94720, USA

[3]Department of Physics and Nanotechnology, SRM Institute of Science and Technology, Kattankulathur, Chennai, Tamil Nadu, India

[4]Molecular Foundry, Lawrence Berkeley National Laboratory, Berkeley, CA 94720, USA

([*kaushalya@lbl.gov](*kaushalya@lbl.gov))

## 1. Sample preparation

Figure S1 shows the depth profile of common elements in an as-received SOI wafer using secondary ion mass spectroscopy (SIMS). Residual hydrocarbons such as $^{12}$C, H, and O are likely from ambient contamination during storage. Prior to ion implantation, the $^{13}$C concentration is below the sensitivity limits of the SIMS measurement (~$10^{16}$ $^{13}$C/cm$^2$). It can be noted that even in the as-received SOI wafer, there are already enough $^{12}$C, H, and O atoms present to write common quantum emitters such as G centers (refer to section 3 for all-optical direct writing of W and G centers on as-received SOI sample without the need for any additional ion implantation or rapid thermal annealing).

SOI wafers were then implanted with carbon ions ($^{13}$C) by Kroko Implants. Prior to ion implantation, Stopping and Range of Ions in Matter (SRIM) simulations were performed to select implantation conditions for the desired C ion implantation depth in the Si device layer. We have targeted the implantation depth to be about in the center of the 220 nm thin Si device layer which is optimized for integrating the quantum emitters into photonic cavities [1]. As per the feedback received from the Stopping and Range of Ions in Matter (SRIM) simulation, ion energy of 38 keV at 7 degrees angle of incidence, provides a well-centered profile of C ions in the Si device layer.

These C-implanted substrates were then subjected to rapid thermal annealing under forming gas ($H_2$:10%, $N_2$:90%) ambiance at 800 °C, for 120 seconds. The stepwise temperature profile for the annealing process representing the time for each step is shown in figure S1 b. We have verified the depth-resolved concentration profile for various common elements (i.e., $^{13}C$, $^{12}C$, H, O) in the pre-processed SOI wafer using SIMS as shown in figure S1 c.

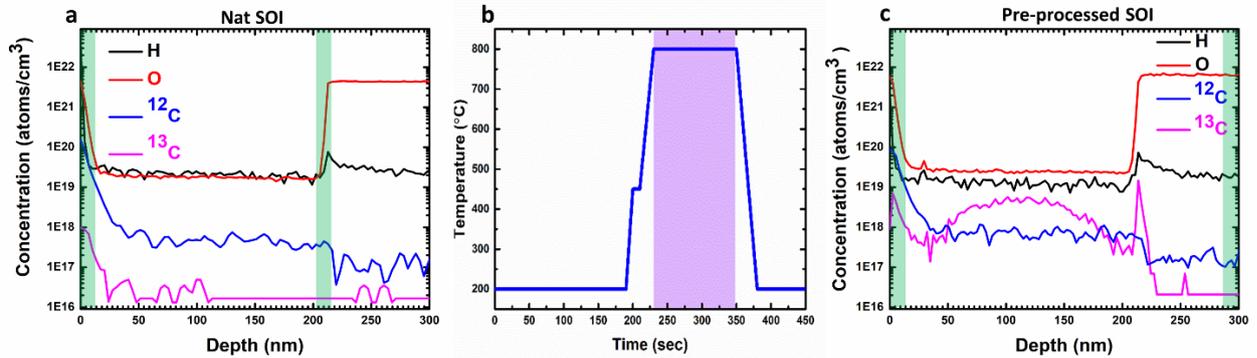

**Figure S1 SIMS depth profile of common elements in SOI device layers before and after ion implantation and rapid thermal annealing. a. SIMS profile representing the concentration of H, O, $^{12}C$, $^{13}C$ in the as-received SOI. b. Process flow for the rapid thermal annealing under forming gas ambiance (H$_2$:10%, N$_2$:90%) at 800 ºC for 120 seconds to form pre-processed SOI samples. c. SIMS profile representing the concentration of H, O, $^{12}C$, and $^{13}C$ in the pre-processed SOI. SIMS data near the surface and the silicon-SiO$_2$ box interface are highlighted in green and show transients to local changes of the relative ionization probabilities.**

Figure S1 c shows SIMS profiles for the pre-processed SOI. As simulated with SRIM, an ion energy of 38 keV of $^{13}C$ ion was used to target the center of the 220 nm thin Si device layer. Indeed, a nice $^{13}C$ profile centered around the desired depth was also observed with SIMS as shown in figure s1 c. The SIMS profile shown in figure s1 c was measured after the rapid thermal annealing of $^{13}C$ implanted SOI.

## 2. $C_i$ center formation in SOI with ion implantation and thermal annealing under forming gas

Figure S2 a show the PL spectrum measured at 6 K, representing the ZPL corresponding to the $C_i$ center. Here, the emission peaks at 1415.4 nm, 1441.7 nm, and 1450.8 nm are associated with different charge states of the $C_i$ center containing $^{13}C$ that was introduced during implantation. A slight shift in the previously reported ZPL associated with the neutral state of the $C_i$ center (i.e., 1448 nm) with that of the one obtained in this study at 1450.8 nm can be attributed to strain effects [2] and the presence of hydrogen in bright Ci centers. Figure S2 b shows the dependence of the excitation power on the PL peak intensity of the ZPLs (1415.4, 1441.7, 1450.8 nm).

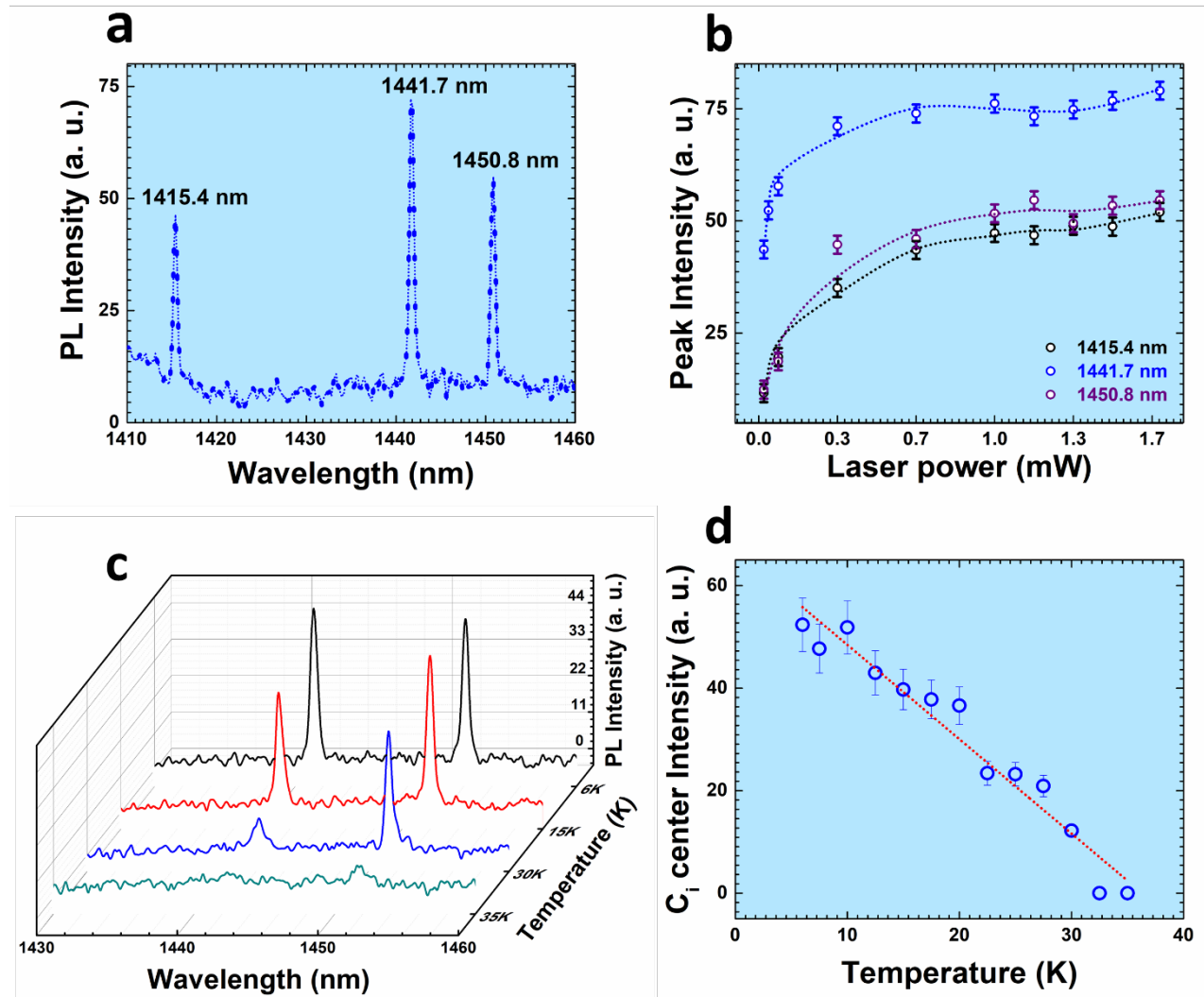

**Figure S2 Bright $C_i$ center realization in SOI with C ion implantation followed by forming gas annealing. a.** PL spectra obtained at 6 K with continuous 532 nm laser excitation representing the zero-phonon line (1415.4, 1441.7, 1450.8 nm) corresponding to the $C_i$ center (these peaks can be associated with the different charged state of the $C_i$ center). **b.** Peak amplitude variation as a function of excitation laser power shows a ~1:2 intensity ratio between the 1441.7 nm and the two other peaks (1415.4 and 1450.8 nm). **c-d.** Temperature response of the emission peak at ~1453 nm and the corresponding PL spectrum shows the robustness and evolution in the $C_i$ center emission from 6-37.5 K.

Depending on the measurement position on the sample, a distribution in the ZPLs was also observed (see supplementary information section 3 for details on the statistics). The intensity variation follows a power function and saturates after a threshold excitation power of ~ 0.7 mW. A ratio of ~1:2 was observed between the intensities of the 1441.7 nm peak and other emission peaks, which is also dependent on the measurement location.

The temperature dependence of PL emission was also monitored to understand the stability and robustness of the emitter. Figure S2 c shows the peak intensity of the $C_i$ centers as a function of temperature (the measurement shown in Figure S2 c was undertaken at a different location on the sample from the one shown in Figure S2 b. A sublinear dependence between the PL peak intensity and the temperature was observed from 6-37.5 K, with PL intensity decreasing with temperature due to increased phononic and non-radiative contributions[3]. PL emission peak intensity corresponding to different temperatures are shown in Figure S2 d. Similar measurements were performed during several temperature cycles and the emission from $C_i$ centers was found to be unaffected reflecting their robustness and stability.

## 3. $C_i$ center statistics on ZPL and high-resolution spectra for linewidth estimation

As discussed in the main manuscript, emission peaks corresponding to different charge states of the $C_i$ center have ZPLs scattered within a few nm from the literature value related to the neutral state of the $C_i$ center (i.e., 1448 nm). Figure S3 a show the PL spectra representing different charge states associated with $C_i$ centers measured at 20 different locations distributed in the area of ~5 mm$^2$ on the pre-processed SOI sample. Figure S3 b shows an extremely narrow linewidth of ~0.03 nm measured with the highest grating (1200 g/mm, having a resolution of ~0.03 nm) indicating the actual linewidth to be even smaller than the one measured.

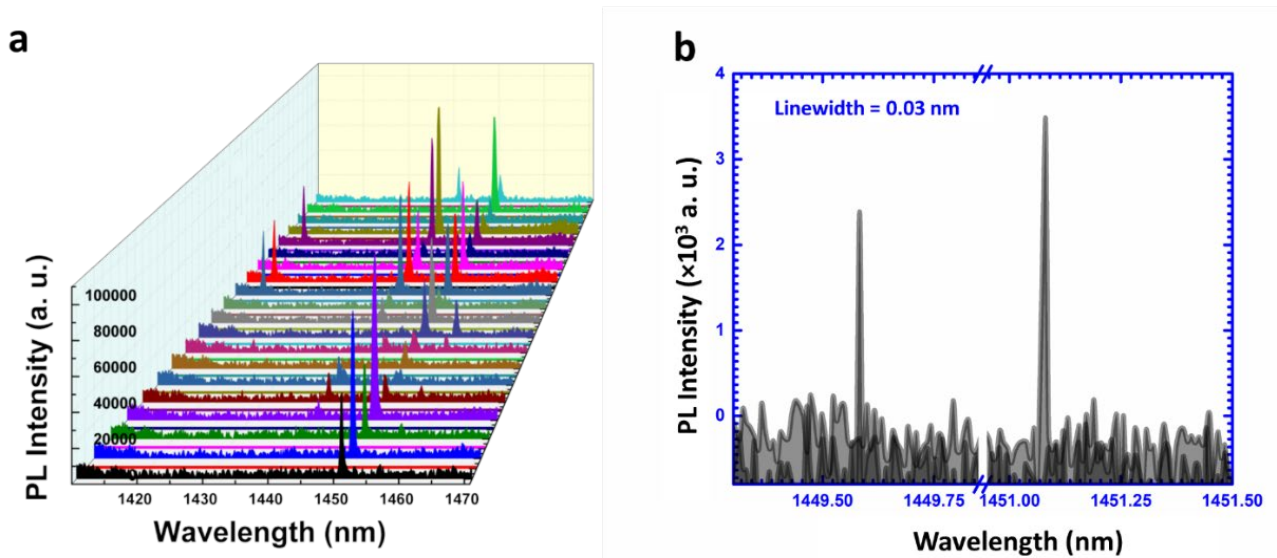

**Figure S3 Distribution in the zero-phonon lines (ZPLs) associated with different charge states of the $C_i$ center and high-resolution PL spectra for linewidth estimation. a.** PL spectra representing different charge states associated with $C_i$ centers measured at 20 different locations distributed in the area of ~5 mm² on the pre-processed SOI. **b.** The High-resolution PL spectrum for $C_i$ centers was measured with 1200 g/mm grating. Linewidth of ~0.03 nm is limited by the resolution of the spectrometer with the highest grating.

## 4. Fs laser irradiation for deterministic writing and erasing of G and $C_i$ centers

As described in the manuscript, a single fs pulse irradiation with varied laser fluences was employed to write and erase various quantum emitters in particular G and $C_i$ centers. Figure S4 shows the PL spectra corresponding to the above-mentioned emitters after fs irradiation at different laser fluences. Here an evolution in the PL emission can be observed for both emitters. Figure S4 shows the PL peak intensity from both emitters.

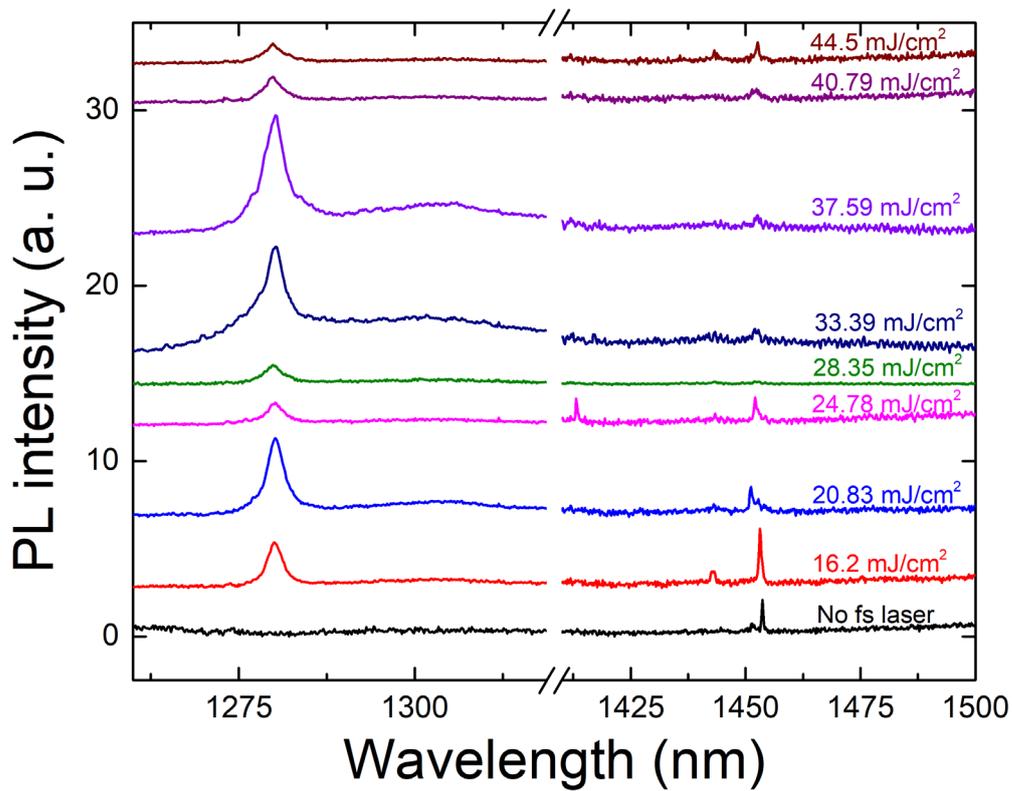

**Figure S4 PL measurements for G and $C_i$ centers before and after fs irradiation in pre-processed SOI. a. PL spectra starting from no fs irradiation (showing only $C_i$ center emission) to single pulse irradiation at varied fluences (showing the evolution in the G and $C_i$ center emission resulting from fs laser driven writing and erasing).**

## 5 First-principles calculations

### 5.1. Charge state stability of $C_i$ centers in silicon

The formation energies of various charge states of the Ci-center in silicon were computed as a function of the position of the Fermi level in the gap, using the Spinney package[4], with and without the finite size correction scheme of Kumagai and Oba[5].

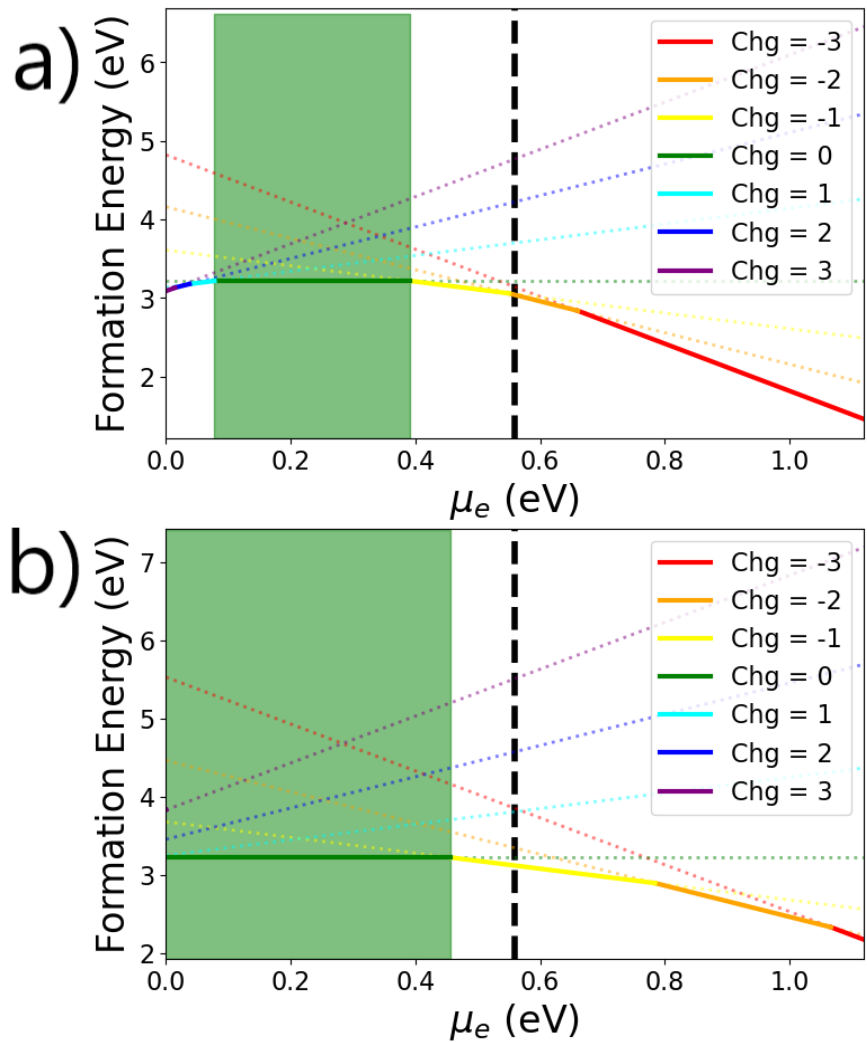

**Figure S5.1 Formation energy vs. Fermi level position in the silicon band gap for the $C_i$ center.** Formation energies for charge states ranging from -3 to +3 as a function of the Fermi level are plotted as dashed lines, with a solid line showing the most stable charge state at each level of doping. The black vertical line shows the position of the Fermi level for neutral, undoped silicon. The shaded green region denotes the region where the neutral charge state is the most stable. Results are shown without (a), and with (b) a finite size correction due to periodic boundary conditions.

**5.2 Hydrogen-modified structures of the G-center**

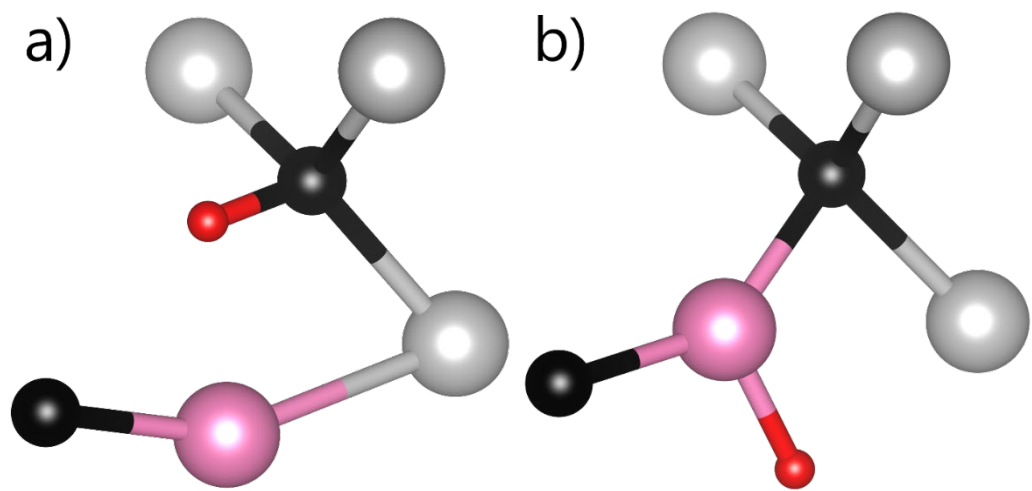

**Figure S5.2 Two structures of the G center B-form augmented by interstitial hydrogen. The possible structures of the G center Type-B modified by hydrogen are shown, with the hydrogen bonded to (a) the substitutional carbon, or (b) the interstitial silicon. Atoms are colored to denote silicon (gray), carbon (black), self-interstitial silicon (pink), and interstitial hydrogen (red).**

# References

1. Redjem, W. *et al.* Single artificial atoms in silicon emitting at telecom wavelengths. *Nature Electronics 2020 3:12* **3**, 738–743 (2020).
2. Liu, W. *et al.* Quantum emitter formation dynamics and probing of radiation induced atomic disorder in silicon. (2023). *Phys. Rev. Applied* (in press), *arXiv: 2302.05814*
3. Beaufils, C. *et al.* Optical properties of an ensemble of G-centers in silicon. *Phys Rev B* **97**, 035303 (2018).
4. Arrigoni, M. & Madsen, G. K. H. Spinney: Post-processing of first-principles calculations of point defects in semiconductors with Python. *Comput Phys Commun* **264**, 107946 (2021).
5. Kumagai, Y. & Oba, F. Electrostatics-based finite-size corrections for first-principles point defect calculations. *Phys Rev B Condens Matter Mater Phys* **89**, 195205 (2014).